\documentclass[preprint*]{JHEP3}

\usepackage{epsfig,multicol}
\usepackage{amsmath}
\usepackage{epstopdf}
\usepackage{placeins}
\newcommand\fverb{\setbox\pippobox=\hbox\bgroup\verb}
\newcommand\fverbdo{\egroup\medskip\noindent%
\fbox{\unhbox\pippobox}\ }
\newcommand\fverbit{\egroup\item[\fbox{\unhbox\pippobox}]}
\newbox\pippobox

\title{Modified spontaneous symmetry breaking pattern
by brane-bulk interaction terms}

\author{Francesco Coradeschi\\
Dipartimento di Fisica, Universit\`a di Firenze,  I-50019 Sesto
F., Italy\\
I.N.F.N.,
Sezione di Firenze,  I-50019 Sesto F., Italy\\
E-mail: \email{coradeschi@fi.infn.it}}

\author{Stefania De Curtis\\
I.N.F.N.,
Sezione di Firenze,  I-50019 Sesto F., Italy\\
E-mail: \email{decurtis@fi.infn.it}}

\author{Daniele Dominici\\
Dipartimento di Fisica, Universit\`a di Firenze,  I-50019 Sesto
F., Italy\\
I.N.F.N.,
Sezione di Firenze,  I-50019 Sesto F., Italy\\
E-mail: \email{dominici@fi.infn.it}}

\author{Jos\'e R. Pelaez\\
Dept. de F\'{\i}sica Te\'orica II,
  Univ. Complutense, 28040 Madrid, Spain\\
  E-mail: \email{jrpelaez@fis.ucm.es}}

\received{}
\revised{}
\accepted{}             

\preprint{\hepth{******}}  

\abstract{
We show how translational invariance can be broken by the  vacuum 
that drives the spontaneous symmetry breaking of extra-dimensional extensions
of the Standard Model, when delta-like
interactions between brane and bulk scalar fields are present.
We explicitly build some examples of vacuum configurations, which  induce the spontaneous symmetry breaking, and have non trivial profile in the extra coordinate.
}


\begin{document}
\newcommand{\NP}[1]{ Nucl.\ Phys.\ {#1}}
\newcommand{\ZP}[1]{ Z.\ Phys.\ {#1}}
\newcommand{\RMP}[1]{ Rev.\ of Mod.\ Phys.\ {#1}}
\newcommand{\PL}[1]{ Phys.\ Lett.\ { #1}}
\newcommand{\NC}[1]{ Nuovo Cimento {#1}}
\newcommand{\AN}[1]{ Ann. Phys. {#1}}
\newcommand{\PRep}[1]{Phys.\ Rep.\ {#1}}
\newcommand{\PR}[1]{Phys.\ Rev.\ { #1}}
\newcommand{\PRL}[1]{ Phys.\ Rev.\ Lett.\ { #1}}
\newcommand{\MPL}[1]{ Mod.\ Phys.\ Lett.\ {#1}}
\newcommand{\IJmp}[1]{{\em Int.\ J.\ Mod.\ Phys.\ }{\bf #1}}
\newcommand{\Od}{O}
\newcommand{\qfint}{\int \frac{d^4 q}{(2\pi)^4}}
\newcommand{\qtint}{\int \frac{d^3 \vec{q}}{(2\pi)^3}}
\newcommand{\qmixv}{(q_0,\vec{q},\tau)}
\newcommand{\qmixw}{(q_0,\omega_q,\tau)}
\newcommand{\tr}{\mbox{tr}}
\newcommand{\Tr}{\mbox{Tr}}
\newcommand{\Dim}{\mbox{dim}}
\newcommand{\im}{\mbox{Im}\,}
\newcommand{\re}{\mbox{Re}\,}
\newcommand{\sgn}{\mbox{sgn}}
\newcommand{\diag}{\mbox{diag}}
\newcommand{\hpisqt}{h_\pi^2 (t)}
\newcommand{\fpitsq}{f_\pi^2 (t)}
\newcommand{\ftsq}{f^2(t)}
\newcommand{\fzerosq}{f^2(0)}
\newcommand{\fdot}{\dot f(t)}
\newcommand{\fddot}{\ddot f(t)}
\newcommand{\tti}{\tilde t}
\newcommand{\lagf}{{\cal L}^{(4)}}
\newcommand{\intc}{\int_C dt \int d^3 \vec{x}}
\newcommand{\intcc}{\int_C \! d^4x}
\newcommand{\vxt}{(\vec{x},t)}
\newcommand{\bea}{\begin{eqnarray}}
\newcommand{\eea}{\end{eqnarray}}
\newcommand{\be}{\begin{equation}}
\newcommand{\ee}{\end{equation}}
\def\bc{\begin{center}}
\def\ec{\end{center}}
\newcommand{\ar}{\arrowvert}
\newcommand{\ra}{\rangle}
\newcommand{\la}{\langle}
\newcommand{\da}{^\dagger}
\newcommand{\ov}{\overline}
\newcommand{\cd}{\! \cdot \!}
\newcommand{\tad}{F_\beta}
\newcommand{\f}{\frac}
\newcommand{\dd}{\displaystyle}
\newcommand{\lr}[1]{{ \left( \, #1 \, \right) }}
\newcommand{\gs}{g'_5 \, \! ^2}
\newcommand{\s}{\sigma}
\newcommand{\sbar}{\bar\sigma}
\newcommand{\smu}{\sigma^\mu}
\newcommand{\sbarmu}{\bar\sigma^\mu}

\newcommand{\nn}{\nonumber}
\def\f{\frac}

\def\dd{\displaystyle}
\def\nn{\nonumber}
\def\sgn{{\rm sgn}}
\def\ad{\dot{\alpha}}
\def\ov{\overline}
\def\cP{{P}}\def\vl{{v_\lambda}}
\def\vl{{v_\lambda}}
\def\v{{v}}
\def\bPsi{{\bf \Psi}}
\def\hbPsi{{\bf \hat\Psi}}
\def\hbPsim{{{\bf \hat \Psi^M}}}

\newcommand{\derbar}{\not{\!\partial}}

\newcommand{\ba}{\begin{eqnarray}}
\newcommand{\ea}{\end{eqnarray}}


\section{Introduction}

The intense  activity over the recent years  in extra dimensional models has renewed the
interest in the study of the vacuum configurations of such theories.
Vacuum solutions with a non trivial
behaviour in the extra coordinate have been investigated in simple five
dimensional models
where particles can be confined to a brane or live in the bulk,
in particular to understand chirality properties or fermion masses and
mixings,
\cite{ArkaniHamed:1999dc,Georgi:2000wb,Kaplan:2001ga}, or
just to study the existence and stability of non trivial scalar
configurations in simple $\lambda\phi^4$ theories on the circle
or the orbifold \cite{Manton:1988az,Grzadkowski:2004mg,George:2006gk,Davies:2007xr,Toharia:2007xe},
extending the pioneering paper on field localization in extra dimensions
by Rubakov and Shaposhnikov \cite{Rubakov:1983bb}.
The aim of this paper is to study the modification of the naive vacuum configuration 
in presence of delta-like interactions between brane and bulk fields.
Brane terms are always generated by radiative corrections, even in the absence of tree level
brane couplings \cite{Georgi:2000wb}. Note that the effect of brane kinetic terms 
has been investigated for scalar, fermion and  gauge theories in \cite{Georgi:2000wb,Carena:2002me,Davoudiasl:2002ua,
delAguila:2003bh,delAguila:2003gv,delAguila:2003gu},
whereas here we will study interaction terms.

For simplicity we will illustrate these effects in a simple two-Higgs doublet model in five dimensions,
assuming one Higgs in the bulk and the second
one on the brane. Models of this type have been considered mainly
from the phenomenological point of view as the simplest extensions of the Standard Model (SM)
in five dimensions without supersymmetry
\cite{Masip:1999mk,Casalbuoni:1999ns,Delgado:1999sv,Rizzo:1999br,Muck:2001yv}.
In the analysis of these
models usually one assumes the existence of a constant vacuum solution
for the bulk field, which does not depend on the extra coordinate,
 without discussing whether the two-Higgs potential
admits such a solution. In a previous paper
\cite{DeCurtis:2002nd} we have already noticed
that a constant solution in general does not exist, unless
a particular relation among the quadrilinear couplings of the
bulk and brane Higgs
 potential is satisfied. In this paper we provide
analytic expressions for vacuum solutions and we build explicit examples
with non trivial profiles corresponding to configurations
which are absolute minima of the energy density.

In Section 2 we review the $SU(2)_L\times U(1)_Y$ two-Higgs model
in five dimensions, with the field  $\Phi_1$ propagating in the bulk
and the field $\Phi_2$ localized on the brane at $y=0$.
In Section 3 we analyze the most general solutions, constant on the brane, of the equations of
motion of the scalar fields in terms of Jacobi elliptic functions. In Section 4
we explicitly build some examples where these solutions have
lower energy than the trivial $\Phi_1=\Phi_2=0$ solution,
showing that, indeed, non trivial vacuum configurations exist.
These solutions lead to a spontaneous symmetry breaking with
a pattern which is non standard, since the vacuum is not translationally
invariant in the extra coordinate, and the vacuum expectation value of the bulk scalar
field is not related to the Lagrangian parameters in the usual manner.

\section{Delta-like interactions between brane and bulk fields}
\label{section2}
For illustration, let us consider a very simple 5D scalar model 
with an action containing both bulk and brane  terms as follows:
\ba
&&S=\int^b_a dy\int d^4x \left\{ {\cal L}^{(5)}+{\cal L}^{(4)}) \right\}\,,\\
&&{\cal L}^{(5)}=\partial_M\Phi_1\da \partial^M\Phi_1-V^{(5)}(\Phi_1)\,,\\
&&{\cal L}^{(4)}=\delta(y)\Big[\partial_\mu\Phi_2\da \partial^\mu\Phi_2-V^{(4)}
(\Phi_1,\Phi_2)\Big]\,,
\ea
where $M=\mu,5$ and $a<0<b$. Note that $\Phi_1$ has
energy dimension 3/2, whereas $\Phi_2$ has dimension 1.
There could be some other fields, but, for the following discussion,
only $\Phi_1$ and $\Phi_2$ are relevant.
In order to identify the vacuum state, we need to solve the equations of motion
\ba
&&(-\partial_y^2+\Box)\Phi_1=\frac{\delta V^{(5)}}{\delta \Phi_1}
+\delta(y)\frac{\delta V^{(4)}}{\delta \Phi_1}\,,\\
&&\delta(y)\Box\Phi_2=\delta(y)\frac{\delta V^{(4)}}{\delta \Phi_2}\,,\\
&&\sum_{\alpha=\Phi_1,\Phi_1\da}\int d^4x \Big[\Big(
\frac{\delta {\cal L}^{(5)}}{\delta\partial_y\Phi_\alpha}\delta\Phi_1\Big)_{y=b}
-\Big(
\frac{\delta {\cal L}^{(5)}}{\delta\partial_y\Phi_\alpha}\delta\Phi_1\Big)_{y=a}\Big]=0\,.
\label{boundaryterms}
\ea
The last term comes from the boundary conditions, and could
also give rise to contributions that can be recast in terms of $\delta(y-a)$ and $\delta(y-b)$ functions
and are thus similar to those that we will consider next.
This said, and for simplicity, we will choose periodic boundary conditions so 
that Eq. \eqref{boundaryterms} is automatically satisfied. If it was not the case, one should repeat for this boundary term the same analysis we will follow below for the $\delta(y)$ term.

The vacuum manifold corresponds to those solutions of the above equations of motion
with minimum energy. Customarily, one considers constant solutions, i.e.,
$\Phi_1=v_1, \Phi_2=v_2$,
so that the vacuum manifold corresponds to the minima of the potential,
and, in particular, $\delta V^{(5)}/\delta \Phi_1=0$ and
$\delta V^{(4)}/\delta \Phi_i=0$ with $i=1,2$.
However, we will show here that the presence of delta-like interactions between brane and bulk fields modifies the vacuum manifold in such a way that static field configurations are not
allowed any more. We will show that this effect is non-perturbative and that even an
infinitesimal value of such a coupling could avoid the presence
of the naively expected pattern of spontaneous symmetry breaking on the brane.

In order to illustrate these effects we will concentrate on a model widely used
in the literature, although our considerations
are applicable to more general solutions of the kind described above
(and probably involving other kind of fields like fermions,
or more complicated interaction
terms, as long as brane-bulk interactions are present).

\subsection{An example within 5D extensions of the Standard Model}
We consider a minimal 5D extension
of the SM with two scalar fields. For the moment it is irrelevant
whether the compactification is done on the $[-\pi R,\pi R]$ circle 
with periodic boundary conditions or in an orbifold
$S^1/Z_2$, of length $\pi R$, since we are only
interested in vacuum configurations. Of course, for oscillations around
the vacuum the orbifold would lead to 
fields with definite $y$-parity.

In this simple model the $SU(2)_L$ and $U(1)_Y$
gauge fields and the Higgs field $\Phi_1$   propagate in the bulk
while  the Higgs field $\Phi_2$ lives on the brane at $y=0$. The
Lagrangian of the gauge Higgs sector is given by (see
\cite{Muck:2001yv}
 for a review)
\ba \int_{-\pi R}^{\pi R} dy\int
dx\,{\cal L}(x,y) &=&\int_{-\pi R}^{\pi R} dy\int dx\,\Big\{ - \frac
1 4 B_{MN}B^{MN} - \frac 1 4 F^a_{MN}F^{aMN}+{\cal L}_{GF}(x,y)
\nonumber\\
&+&
(D_M \Phi_1)^\dagger (D^M \Phi_1)+\delta(y)
(D_\mu \Phi_2)^\dagger (D^\mu \Phi_2)- V(\Phi_1,\Phi_2)
\Big\},
\label{kinlagrangian}
\ea
$B_{MN}$, $F_{MN}^a$ are
the $U(1)_Y$ and $SU(2)_L$ field strengths and
 $a$ is the $SU(2)_L$ index.
The covariant derivative is defined as
$D_M=\partial_M - i g_5 A^a_M\tau^a/2 - i g_5'B_M/2$.
For simplicity we will consider a Higgs potential symmetric under
the discrete symmetry $\Phi_2\rightarrow -\Phi_2$, which is given by
\begin{eqnarray}
\label{scalarpotential}
V(\Phi_1,\Phi_2) \!\!\!&=&\!\! \mu_1^2 \, ( \Phi_1\da \Phi_1 ) \, + \,
\lambda_1 \, ( \Phi_1\da \Phi_1 )^2 \,
                            + \, \delta(y) \, \Big[\,
\frac{1}{2}\,\mu_2^2 \, ( \Phi_2\da \Phi_2 ) \,
+ \,
\frac{1}{2}\,\lambda_2 \, ( \Phi_2\da \Phi_2 )^2 \nonumber\\
&+&\!\!\!
\frac{1}{2}\,\lambda_3 \, ( \Phi_1\da \Phi_1 ) ( \Phi_2\da \Phi_2 ) \,
   + \, \frac{1}{2}\,\lambda_4
\, ( \Phi_1\da \Phi_2 ) ( \Phi_2\da \Phi_1 )\,
   + \, \lambda_5\, ( \Phi_1\da \Phi_2 )^2\,
+ \, {\rm h.c.}\, \Big],
\end{eqnarray}
where the dimensionalities
of the couplings are: 1 for $\mu_1$ and $\mu_2$,
 -1 for
$\lambda_1,\lambda_3,\lambda_4$ and
$\lambda_5$,
whereas $\lambda_2$ is dimensionless.
The vacuum state manifold corresponds to configurations 
which are both energy minima and solutions of the following equations of motion:
\ba
(-\partial_y^2+\Box)\Phi_1&=&\mu_1^2\Phi_1+2\lambda_1 \, (\Phi_1\da \Phi_1 ) \Phi_1
\label{EOM1}\\
&+&\delta(y)\left[\lambda_3 \,
\Phi_1  ( \Phi_2\da \Phi_2 ) + \lambda_4
\,\Phi_2 ( \Phi_2\da \Phi_1 )\, + 2 \, \lambda_5\, ( \Phi_1\da \Phi_2 )\Phi_2 \,\right] \nonumber\,,\\
\delta(y)\Box\Phi_2&=&\delta(y)\Big[\mu_2^2\Phi_2+
2\lambda_2 \, (\Phi_2\da \Phi_2 ) \Phi_2 +
\lambda_3 \, ( \Phi_1\da \Phi_1 ) \Phi_2 \nonumber\\
&&
\qquad+\lambda_4 \, ( \Phi_2\da \Phi_1 ) \Phi_1 + \, 2\lambda_5\, ( \Phi_1\da \Phi_2 ) \Phi_1\Big]\,.
\label{EOM2}
\ea

However, one could naively think, and it is customarily assumed 
\cite{Masip:1999mk,Casalbuoni:1999ns,Delgado:1999sv,Rizzo:1999br,Muck:2001yv}, that 
the extrema of the potential correspond to constant configurations
$\Phi_1=(0,v_1/\sqrt{4\pi R})$,
$\Phi_2=(0,v_2/\sqrt{2})$.
Let us note, however, that if we substitute such constant solutions into
the equations of motion above, we find
\ba
0&=&v_1\left(\mu_1^2+2\lambda_1\frac{v_1^2}{4\pi R}\right)\,,\\
0&=&v_1 \,v_2^2\,\left(\lambda_3+\lambda_4+2\lambda_5\right)\label{funnyconstraint}\,,\\
0&=&v_2\left(\mu_2^2+\lambda_2 v_2^2 + \frac{v_1^2}{4\pi R}(\lambda_3+\lambda_4+2\lambda_5)\right)\,.
\ea
If the trivial solutions $v_1=v_2=0$ correspond to a minimum
we have a trivial vacuum configuration and no
spontaneous symmetry breaking.
When implementing a spontaneous symmetry breaking
one customarily builds the Lagrangian in such a way that $\mu_1^2<0$, $\mu_2^2<0$
and thus 
$v_1\neq0$ and $\v_2\neq0$ correspond to the minimum.
But, due to Eq.\eqref{funnyconstraint}, this can only happen if 
$\lambda_3+\lambda_4+2\lambda_5=0$.
This may come as a surprise since these constants
parametrize the interaction of brane and bulk fields
and are, in principle, independent.
Thus, {\it even the tiniest value of an interaction
with $\lambda_3+\lambda_4+2\lambda_5\neq0$
destroys the usual ansatz of a translationally invariant
vacuum state in the $y$ direction}.

In a previous work \cite{DeCurtis:2002nd}, for simplicity we required $\lambda_3+\lambda_4+2\lambda_5=0$,
which ensures that the minimum of the potential corresponds to the usual ansatz.
In this way, the Higgs fields are expanded in the standard form
\be
\Phi_1(x,y)=\left (
\begin{array}{c}
 \frac{i}{\sqrt{2}}(\omega^1-i\omega^2)\\
\dd { \frac{1}{\sqrt{2}}(\f {v_1}{\sqrt{2\pi R}}+ h_1-i \omega^3) }
\\
\end{array}
\right), \,\,\,
\Phi_2(x)=\left (
\begin{array}{c}
  \frac{i}{\sqrt{2}}(\pi^1-i\pi^2)\\
\dd { \frac{1}{\sqrt{2}}( {v_2}+ h_2-i \pi^3) }
\\
\end{array}
\right),
\ee
where $v_1\equiv \sqrt{-2 \pi R
\mu_1^2/\lambda_1}$ and $v_2 \equiv \sqrt{-\mu_2^2/\lambda_2}$ are the vacuum expectation values of the scalar fields and $v^2=v_1^2+v_2^2=(\sqrt{2} G_F)^{-1}$. 
Let us remark that we assume $\lambda_1>0$,
$\lambda_2>0$ and $\lambda_3>-2\sqrt{2 \pi R \lambda_1 \lambda_2}$, otherwise the potential will not be bounded from below.

In the literature this ``constant ansatz'' is sometimes assumed \cite{Muck:2001yv} without 
noting that
the relation $\lambda_3+\lambda_4+2\lambda_5=0$ is required,
which thus limits the generality of the approach.
We will see that, for certain choices of parameters, the assumption that the vacuum 
state is independent of $y$
might still be a good approximation, although, as we will show, the vacuum expectation
value of $\Phi_1$ could be rather different from that of the constant case. 
Moreover, one could wonder what happens in the general case
when $\lambda_3+\lambda_4+2\lambda_5\neq0$ at tree level or 
if such a term was generated at higher orders 
from the different renormalization of the 
$\lambda_3$, $\lambda_4$ and $\lambda_5$ parameters.
We will see that, by including such a term, the spatial invariance in 
the fifth dimension $y$ is broken and 
non trivial vacuum configurations should be obtained
from solutions of Eqs.(\ref{EOM1}) and (\ref{EOM2}).

\section{Static solutions of the equations of motion}
\label{section3}
\def\eps{{\epsilon}}
\def\mub{{\bar\mu}}
\def\lambdab{{\bar\lambda}}

Following the previous discussion, in this Section we will
first  search for  solutions of the equations of motion
that could play the role of the true vacuum. In Section \ref{examples}, we will
study whether these solutions have a lower energy
than the trivial vacuum so that they can trigger a spontaneous symmetry breaking. 
In particular
we will look here for  solutions that do not
depend on the 4D space-time coordinates $x$, but still
have a dependence on $y$.

For the sake of simplicity, and because we just want
to illustrate the effects due to the presence of a $\delta(y)$ term, 
we will study the $\lambda_4=\lambda_5=0, \lambda_3\neq0$ case,
since we can then recast the static vacuum solutions as
\begin{equation}
\langle \Phi_1(x,y) \rangle=\left (
\begin{array}{c}
 0\\
\varphi_1(y) 
\\
\end{array}
\right), \,\,\,
\langle \Phi_2(x) \rangle=\left (
\begin{array}{c}
  0\\
\varphi_2
\\
\end{array}
\right),
\end{equation}
where $\varphi_1(y)$ is a real-valued field, and $\varphi_2$ a real constant.
Therefore, the equations of motion, Eqs.(\ref{EOM1}) and (\ref{EOM2})
for non-trivial vacuum solutions 
in this model,  are reduced to
\ba
&&\partial_y^2\varphi_1(y)-\varphi_1(y)
\left[\mu_1^2+2\lambda_1 \varphi_1(y)^2 +\delta(y)\,\lambda_3  \varphi_2^2 \right]=0\,,
\label{e:0}\\
&&\delta(y)\,\varphi_2\left[\mu_2^2+ 2\lambda_2
\, \varphi_2^2  + \lambda_3 \, \varphi_1(y)^2\right]=0\,.
\label{e:1}
\ea
The above solutions have an associated energy density per unit volume:
\begin{equation}
\label{energy}
\!\mathcal{H} =\!\! \int_{-\pi R}^{\pi R} \!\!\!\!\! dy \left[(\partial_y \varphi_1(y))^2 + \mu_1^2 \varphi_1(y)^2 + \lambda_1 \varphi_1(y)^4 + \delta(y) \left(\mu_2^2 \varphi_2^2 + \lambda_2 \varphi_2^4 + \lambda_3 \varphi_1(y)^2 \varphi_2^2 \right) \right].
\end{equation}

As usually done, we account for the presence of the $\delta$-function
by solving the $\delta$-less equation
\begin{equation}
  \partial_y^2\varphi_1(y)-\varphi_1(y)\left[\mu_1^2+2\lambda_1 \varphi_1(y)^2\right]=0
        \label{e:b1}
\end{equation}
in the bulk regions $y<0$ and $y>0$ separately, and then connecting 
both pieces using the following boundary conditions
\begin{itemize}
\item continuity in $y=0$:
\begin{equation}
        \varphi_1(0^-) = \varphi_1(0^+) \equiv \varphi_1(0);
        \label{e:b2}
\end{equation}
\item discontinuity of the first derivative in $y=0$ with a gap $\lambda_3 \varphi_2^2 \varphi_1(0)$:
\begin{equation}
\varphi_1'(0^+)-\varphi_1'(0^-) = \lambda_3 \varphi_2^2 \varphi_1(0),
\label{e:b3}
\end{equation}
where by Eq.(\ref{e:1}) we should have
\begin{equation}
        \varphi_2^2 = -\frac{\mu_2^2}{2 \lambda_2} - \frac{\varphi_1(0)^2 \lambda_3}{2 \lambda_2}, \quad {\rm with} \quad  \varphi_2^2  > 0.
        \label{e:b4}
\end{equation}
\end{itemize}

\subsection{Solutions in the bulk}
Let us solve Eq.\eqref{e:b1}. Following \cite{Rubakov:2002fi,Grzadkowski:2004mg} we first multiply both sides by $\partial_y \varphi_1(y)$ and integrate in $y$,  to get
\begin{equation}
\f 1 2 (\partial_y \varphi_1(y))^2-\f 1 2 \mu_1^2 \varphi_1(y)^2-\f{\lambda_1}{2}\varphi_1(y)^4=e_0,
\end{equation}
where $e_0$ is a conserved quantity.
Thus, integrating again
\begin{equation}
y-y_0=\pm\int_{\varphi_1(y_0)}^{\varphi_1(y)} \f{d t}{\sqrt{\mu_1^2 t^2+\lambda_1 t^4+2 e_0}}.
\label{int}
\end{equation}

This integral can be solved analytically in terms of Jacobi elliptic functions
\cite{Stegun:1964,Grad:1964}. Such methods are well known, and thus we only provide the necessary steps to understand our notation. 
In particular, the exact solution depends on the nature of the roots of the polynomial
\begin{equation}
P_4(t) \equiv \lambda_1 t^4 + \mu_1^2 t^2+2 e_0.
\end{equation}
These are given by
\begin{equation}
t^2 = \frac{-\mu_1^2 \pm \sqrt{\mu_1^4-8e_0 \lambda_1}}{2 \lambda_1} \equiv \frac{-\mu_1^2}{2 \lambda_1} \left(1\mp \beta^2 \right),
\end{equation}
with
\begin{equation}
\beta^2=\sqrt{1-\alpha}, \quad \alpha = \frac{8 e_0 \lambda_1}{\mu_1^4}.
\end{equation}

Hence, depending on the values of $\alpha$, we have the following cases:

\begin{itemize}

\item [A)] $\alpha < 0; \ P_4(t)$ has two real 
and two complex solutions. We can therefore make use
of the definition of the Jacobi elliptic ${\rm cn}(x,k^2)$ function:
\begin{equation}
\int_1^x \frac{dt}{\sqrt{(1-t^2)(1-k^2+k^2 t^2)}} = {\rm cn}^{-1}(x,k^2),
\end{equation}
to rewrite Eq.(\ref{int})  as follows:
\begin{equation}
y-y_0= \pm \frac{a}{\sqrt{N}}\int_{\frac{\varphi_1}{a}(y_0)}^{\frac{\varphi_1}{a}(y)} \frac{dt}{\sqrt{(1-t^2)(1-k^2+k^2 t^2)}}. 
\end{equation}
This is achieved by rescaling $t \to at$,  so that
\begin{equation}
P_4(t) \to \lambda_1 a^4 t^4 + a^2 \mu_1^2 t^2 + 2 e_0 
\equiv N (1-t^2)(1-k^2+k^2 t^2),      
\end{equation}
where
\begin{equation}
k^2 = \frac{1}{2}\left(1+\frac{1}{\beta^2}\right),
\ a^2 = \frac{-\mu_1^2}{2 \lambda_1} (1+\beta^2) > 0,
\ N= \frac{- \mu_1^4}{2 \lambda_1} \beta^2(1+\beta^2) < 0.
\end{equation}

In this way we finally get what we will call the ``A type'' solution
\begin{equation}
\varphi_1^A(y) =\pm
\frac{|\mu_1|}{\sqrt{2 \lambda_1}} \sqrt{1+\beta^2} \ {\rm nc}\left(|\mu_1| \beta (y-y_0),\frac{1}{2}(1-\frac{1}{\beta^2})\right),
\label{sol:a}
\end{equation}
where we used the relation ${\rm cn}(i x,k^2)=\dfrac{1}{{\rm cn}(x,1-k^2)} \equiv {\rm nc}(x,1-k^2)$.

\item [B)] $0 \le \alpha \le 1$, that is, $0\le \beta \le 1$; in this case, $P_4(t)$ has four real solutions. 

Again, we rescale $t \to at$; then we can match $P_4(t)$ to
\begin{equation}
P_4(at) \to N (1-t^2)(1-k^2 t^2),
\end{equation}
which leads to a Jacobi elliptic sn$(x,k^2)$ solution
\begin{equation}
\int_0^x \frac{dt}{\sqrt{(1-t^2)(1-k^2 t^2)}}= {\rm sn^{-1}}(x),
\end{equation}
with
\begin{equation}
k^2 = \frac{1-\beta^2}{1+\beta^2},\qquad
a = \frac{|\mu_1|}{\sqrt{2 \lambda_1}} \sqrt{1-\beta^2},
\qquad N= 2 e_0 = \frac{\mu_1^4}{4 \lambda_1} (1- \beta^4) >0 \,,
\end{equation}
thus leading to what we will call ``B1 type'' solution
\begin{equation}
\varphi_1^{B1}(y) = \pm \frac{|\mu_1|}{\sqrt{2 \lambda_1}} \sqrt{1-\beta^2} 
\ {\rm sn}\left(\frac{\vert \mu_1\vert}{\sqrt{2}} \sqrt{1+\beta^2}(y-y_0),\frac{1-\beta^2}{1+\beta^2}\right),
\label{sol:b1}
\end{equation}
which is an oscillating function of $y$ that satisfies
$\varphi_1^{B1}(y_0) = 0$.

But we can also recast $P_4(t)$ as
\begin{equation}
P_4(at) \to N(1-t^2)(t^2-1+k^2),
\end{equation}
which now leads to a Jacobi elliptic dn$(x,k^2)$ solution
\begin{equation}
\int_1^x \frac{dt}{\sqrt{(1-t^2)(t^2-1+k^2)}}={\rm dn^{-1}}(x),
\end{equation}
by identifying,
\begin{equation}
k^2 = \frac{2 \beta^2}{1+\beta^2},\qquad a = \frac{|\mu_1|}{\sqrt{2 \lambda_1}} \sqrt{1+\beta^2}, \qquad
N= \frac{-\mu_1^4}{4 \lambda_1} (1+ \beta^2)^2 <0.
\end{equation}
This is what we will call a ``B2 type'' solution, which does not oscillate.
It satisfies $\varphi_1^{B2}(y_0)/a = 1$, and can be written as
\begin{equation}
\varphi_1^{B2}(y)=\pm \frac{|\mu_1|}{\sqrt{2 \lambda_1}} \sqrt{1+\beta^2} 
\ {\rm dc}\left(   \frac{|\mu_1|}{\sqrt{2}} \sqrt{1+\beta^2} (y-y_0)  ,\frac{1- \beta^2}{1+\beta^2}\right),
\label{sol:b2}
\end{equation}
where we have used the relation ${\rm dn}(i x,k^2)={\rm dc}(x,1-k^2)$
\item [C)] $\alpha > 1$. In this case $\beta^2$ is pure imaginary and $P_4(t)$ has no real solutions. 
We can rewrite Eq. \eqref{int} as:
\begin{equation}
y-y_0=\pm \int_{\tilde{\varphi_1}(y_0)}^{\tilde{\varphi_1}(y)} \f{d\tilde{t}}{\sqrt{(\tilde{t}^2-(1+\sqrt{1-\alpha}))(\tilde{t}^2-(1-\sqrt{1-\alpha}))}},
\end{equation}
where we have made the rescaling:
\begin{equation}
t \to \tilde{t}=\frac{\sqrt{2\lambda_1}}{|\mu_1|} t.
\end{equation}
This integral is not equal to the inverse of a Jacobi elliptic function, as those of the previous cases. However, although in a somewhat more tedious way, it can be solved by using the standard techniques for elliptic integrals \cite{Stegun:1964,Grad:1964}.
The general solution is:
\begin{eqnarray}
\mspace{-42.0mu} \varphi_1^C(y) &=& \frac{|\mu_1|}{\sqrt{2\lambda_1}} 
\sqrt{\frac{1}{2k^2-1}} \times\\
&&\frac{
\textrm{dn}
\left(\frac{|\mu_1|}{\sqrt{2}}\sqrt{\frac{1}{2k^2-1}}\,(y-y_0),k^2\right)
\pm \sqrt{1-k^2} 
\,\,\textrm{sc}
\left(\frac{|\mu_1|}{\sqrt{2}}\sqrt{\frac{1}{2k^2-1}}\,(y-y_0),k^2\right)
}
{
\textrm{dn}
\left(\frac{|\mu_1|}{\sqrt{2}}\sqrt{\frac{1}{2k^2-1}}\,(y-y_0),k^2\right)
\mp \sqrt{1-k^2}
\,\,\textrm{sc}
\left(\frac{|\mu_1|}{\sqrt{2}}\sqrt{\frac{1}{2k^2-1}}\,(y-y_0),k^2\right)
}\,,
\nonumber
\end{eqnarray}
with
\begin{equation}
k^2=\frac{1}{2}\left(1+\frac{1}{\sqrt{\alpha}}\right)\,.
\end{equation}
\end{itemize}

Let us now build the complete solutions of Eq.(\ref{e:b1}) by imposing suitable boundary conditions in $y=0$ and $y=\pi R$.     

\subsection{Matching conditions}

From integration, we initially have four free constants, two
on the left side of the brane $y<0$,
that we call $y_{0L}$ and $\beta_{L}$ ($y_{0L},\alpha_L$ in the case of type C solutions),
and two more on the right side, $y>0$,
called $y_{0R}$ and $\beta_{R}$ (again, $y_{0R},\alpha_R$ for solutions of type C).
This fixes the shape of the function in the intervals,
but, since the fields and their derivatives always
appear squared in the action, there is an overall sign ambiguity,
as it happens in the naive case with $\lambda_3=\lambda_4=\lambda_5=0$
where the vacuum in the bulk is given by either $v_1$ or $-v_1$.

Nevertheless, we are just looking for static minima of the
action, which is symmetric under $y\leftrightarrow-y$. Hence the vacuum states
must be even or odd under $y\leftrightarrow-y$, which implies
$\beta_L=\beta_R\equiv\beta$. Also note that solutions which are antisymmetric under
$y \leftrightarrow -y$ satisfy trivially the boundary condition \eqref{e:b3}; however if we require
the continuity of $\varphi_1(y)$ in $y=\pi R$, its derivative has at least two nodes
(one in the $(0,\pi R)$ region and the other in the $(-\pi R,0)$ one), so it cannot correspond 
to a global minimum of the energy (as we have explicitly checked numerically).
In conclusion we are only interested in \emph{even} solutions and therefore  $y_{0L}=-y_{0R}\equiv y_0$.

Summarizing, apart from the overall sign arbitrariness,
we are left with two constants $\beta, y_0$ that
parametrize the space of possible candidates for vacuum configurations. 

Furthermore our solutions should be of class $C^1$ except 
in $y=0$,  and possibly in $y=\pm \pi R$ where we could impose 
some additional boundary conditions.
At $y=0$ the left and right solutions
should match each other according to Eqs.(\ref{e:b2}), (\ref{e:b3}) and (\ref{e:b4}).
The first one is automatically satisfied for even or odd functions,
as in our case.
If non trivial solutions do exist, then we must have $\mu_2^2<0$, so
\eqref{e:b4} tells us that, for $\lambda_3>0$, $\varphi_1(0)$ is bounded by
$\varphi_1(0)^2<-\mu_2^2/\lambda_3$.
However, Eq.(\ref{e:b3}) gives a relation between the two parameters
$\beta, y_0$, that has to be solved numerically. All in all, 
that leaves us with just one free parameter.
This one can be fixed if we impose an additional boundary condition on $y=\pm \pi R$.
As we will see, the boundary condition could be as simple as requiring
continuity of the first derivative in $\pm \pi R$,
but other choices are possible. Similarly to the terms in Eq.\eqref{boundaryterms}, one could even think
of another delta-like interaction term
localized in a mirror brane in $y=\pm \pi R$.

In summary, by imposing the $y=0$ boundary conditions
in Eqs.(\ref{e:b2}), (\ref{e:b3}) and (\ref{e:b4}),
together with an additional boundary condition on $y=\pm \pi R$,
one has sufficient constraints to fix, up to a global sign,
the complete vacuum configuration in terms of the 
bulk solutions A, B1, B2, C detailed in the previous Section.
In general we found that a given choice of parameters does not allow the existence of
all kind of solutions.
Of course, the trivial solution $v_1= v_2=0$ is always present,
but it will not correspond to the true vacuum
if one of the solutions described above has a lower energy.
This will lead to a spontaneous symmetry breaking
with a pattern that does not correspond to the one
customarily assumed in the literature, since the vacuum 
is not translationally invariant in the $y$ variable.
For some choice of parameters it can also happen that non trivial
solutions cannot be found, so that there is no spontaneous breaking of symmetry.

In the next Section we will show, with explicit
examples, that for certain choices of the parameters, 
the non-trivial configurations do exist and have lower energy densities than the 
trivial one.

\section{Examples of non-trivial vacuum configurations}
\label{examples}

We have seen how the solutions are basically
fixed by the boundary conditions, once we know the Lagrangian parameters.
Let us now remark that in the model we have considered in Section \ref{section2} there
are five independent parameters $\lambda_1, \lambda_2,\lambda_3$
and $\mu_1, \mu_2$. Note that,  in  the realistic case for the usual two-Higgs doublet 
one customarily chooses the parameters with the constraint  
$v_1^2+v_2^2=v^2=(246 \ {\rm GeV})^2$, which fixes 
one of the Lagrangian parameters in terms of the others, and provides the standard mass for
the electroweak gauge bosons once the covariant $SU(2)_L\times U(1)_Y$ derivatives are
considered.

Since we want to illustrate
how the symmetry breaking pattern can be modified 
with non-trivial brane interactions, we will impose a similar constraint.
However, since $\varphi_1(y)$ is not a constant,
we have to look back to the kinetic 
terms of the scalar fields in the
Lagrangian in Eq.(\ref{kinlagrangian}).
Recalling that $D_M=\partial_M - i g_5 A^a_M\tau^a/2 - i g_5'B_M/2$,
we see that the Kaluza-Klein zero modes of the gauge fields
will obtain their masses from the vacuum configuration $\varphi_1(y)$
of the scalar field $\Phi_1$ and the vacuum expectation value 
$\varphi_2$ of $\Phi_2$ through the combination
\be
\label{constraint}
v^2\equiv 4\int_0^{\pi R} \varphi_1(y)^2\,dy + 2 \,\varphi_2^2=
4\int_0^{\pi R} \varphi_1(y)^2\,dy - \frac{\mu_2^2+\lambda_3\,\varphi_1(0)^2}{\lambda_2}\,,
\ee
where we have used Eq.(\ref{e:b4}).
Of course, in the $\lambda_3 \to 0$ limit, we recover the
usual relation $v^2=v_1^2+v_2^2$, but, in our case, the lack 
of translational invariance on $y$ requires an integration
of $\varphi_1(y)^2$ over the compactified fifth dimension. 
Once again, imposing that for the true vacuum
$v^2=(246 \ {\rm GeV})^2$, with $v$ defined in Eq.\eqref{constraint},
fixes one of the Lagrangian parameters 
in terms of the others.

We will show that, depending on the boundary
conditions on $y= \pm \pi R$, we can still find solutions for which the 
usual constant ansatz may be a good approximation, although the vacuum expectation 
value of the $\Phi_1$ field on the $y=0$ brane
might be rather different from $v_1$. In addition, there are
solutions which are change sizably in the extra dimension and should not be approximated 
by a constant value. Both cases will be illustrated with the following examples.

\subsection{Quasi-constant vacuum in the extra dimension}
Let us impose, as a boundary condition, the continuity of the first derivative of $\varphi_1(y)$ in
$y= \pi R$.
The periodicity, moreover, identifies the point $\pi R$ with the point 
$-\pi R$; so, what we require is: $\varphi_1'(\pi R) = \varphi_1'(-\pi R)$. But since $\varphi'_1$
is an odd function, then $\varphi_1'(-\pi R) = - \varphi_1'(\pi R)$ also comes true, so we
conclude that $\varphi_1'(\pi R)=0$, that is, $\pi R$ is a maximum or a 
minimum for $\varphi_1(y)$.
Let us then make a simple choice of parameters:
\begin{equation}
\pi R = (1 {\rm TeV})^{-1}, \quad \mu_1=165 \ {\rm GeV}, \quad \lambda_1= 0.5 \times 2\pi R, 
\quad \lambda_2 = 1, \quad \lambda_3 = 0.85 \times 2 \pi R.
\label{para1}
\end{equation}
Since we require $v=246\,{\rm GeV}$ in Eq.\eqref{constraint}, apart from a global sign,
there is only one continuous solution of Eq. \eqref{e:b3}, that turns out to be of the B1-type, and can be written as follows:
\begin{eqnarray}
\varphi_1^{B1}(y) = + \frac{|\mu_1|}{\sqrt{2 \lambda_1}} \sqrt{1-\beta^2}
\ {\rm sn}\left(\frac{\vert \mu_1\vert }{\sqrt{2}} \sqrt{1+\beta^2}(y-y_0),\frac{1-\beta^2}{1+\beta^2}\right),\quad y>0\,,
\label{sol:b1pos}\\
\varphi_1^{B1}(y) = - \frac{|\mu_1|}{\sqrt{2 \lambda_1}} \sqrt{1-\beta^2} 
\ {\rm sn}\left(\frac{\vert \mu_1\vert }{\sqrt{2}} \sqrt{1+\beta^2}(y+y_0),\frac{1-\beta^2}{1+\beta^2}\right),\quad y<0\,,
\label{sol:b1neg}
\end{eqnarray}
with $\mu_2 \simeq 220$ GeV, $\beta\simeq 0.79$ and $y_0\simeq 0.012$ GeV$^{-1}$. 
Here, for definiteness, we have taken the sign in front of the $y>0$ solution to be positive,
but of course, there is another solution with a global sign difference
that will have the same energy.

\begin{figure}[ht!]
\begin{center}
\epsfxsize = .55\textwidth
\epsfbox{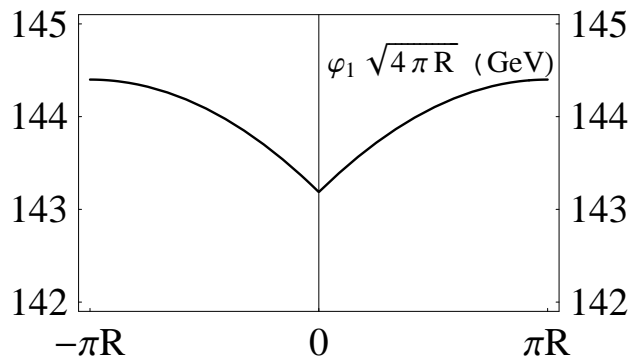}
\end{center}
\caption{\textit{\footnotesize{Vacuum configuration for the choice of parameters of Sect. 4.1. Note that, by taking it as constant, (as it is for $\lambda_3=0$), might be a good approximation, since its variation from $y=0$ to $y=\pi R$ is less than $1\%$. However, $\varphi_1(0)\simeq233\,{\rm GeV}$
 when $\lambda_3=0$, instead of $\varphi_1(0)\simeq143\,{\rm GeV}$ here.}}}
\label{fig:b1}
\end{figure}

The energy density can be calculated using Eq.~(\ref{energy}); we find that it is equal to $-(179\ {\rm GeV})^4$, which is less than the $(0 \ {\rm GeV})^4$ associated with the trivial static solution, thus confirming the fact that we are in presence of a spontaneous symmetry breakdown. Actually, since there are no other solutions, the one we have found, shown in Fig.~\ref{fig:b1} corresponds to a global minimum and can be identified with the true vacuum. As it can be seen from the figure, a constant solution in this case would be an adequate approximation, since the difference between $\varphi_1(0)$ and $\varphi_1(\pi R)$ is less than $1 \%$. However, we should note that the vacuum expectation value of the $\Phi_1$ field on the $y=0$ brane is $\varphi(0)\simeq 139\, {\rm GeV}$, very different from the corresponding $v_1 \simeq 233$ GeV which would be obtained with the parameter choice \eqref{para1} and $\lambda_3 = 0$. This is a $63\%$ decrease that can modify the spectrum of the Kaluza-Klein excitations with respect to the one of the naive ansatz even if the vacuum configuration is almost constant.

\subsection{Sizable violation of translational invariance in the extra dimension}
Let us allow for a discontinuity of the first derivative in $y=\pi R$ assuming for $\varphi'_1(\pi R)$ a given value different from $0$. We make a different choice of parameters:
\begin{equation}
\pi R = (1 {\rm TeV})^{-1}, \quad \mu_1=60 \ {\rm GeV}, \quad \lambda_1= 0.5 \times 2\pi R, 
\quad \lambda_2 = 2, \quad \lambda_3 = 10 \times 2 \pi R.
\label{para2}
\end{equation}
Again, the minimum corresponds to a B1 type solution as in Eqs.~(\ref{sol:b1pos}) and \eqref{sol:b1neg}, 
but with $\mu_2\simeq349$ GeV, $
\beta \simeq 0.1$ and $y_0\simeq 0.15\, {\rm GeV}^{-1}$. The energy density in this case is $\simeq -(245 \,{\rm GeV})^4$, again indicating a spontaneous symmetry breaking.
Incidentally, in this case there is also another solution, of type A, 
but it has a positive energy density and thus it does not correspond to a vacuum state. 

\begin{figure}[ht!]
\begin{center}
\epsfxsize = .55\textwidth
\epsfbox{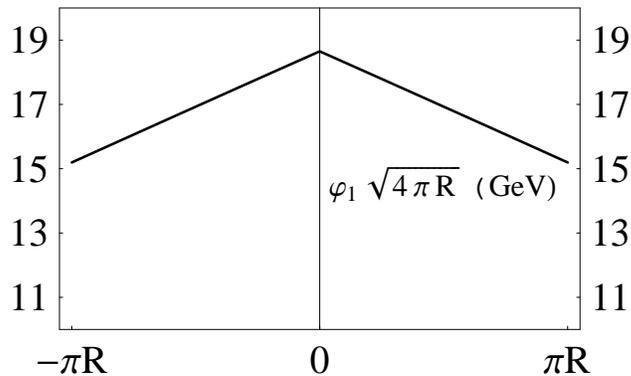}
\end{center}
\caption{\textit{\footnotesize{Vacuum configuration for the choice of parameters of Sect. 4.2. We see that a constant $\varphi_1$ is not
a good approximation: the variation from $y=0$ to $y=\pi R$ is  about $22\%$. The variation of $\varphi_1(0)$ with respect to the non-interacting case is even greater; we would have $\varphi_1(0)\simeq 85\,{\rm GeV}$ for $\lambda_3=0$ (with the other parameters kept constant), while $\varphi_1(0)\simeq 19\,{\rm GeV}$ here.}}}
\label{fig:b1big}
\end{figure}

In Fig. \ref{fig:b1big}, we show the vacuum configuration for the choice of parameters of Eq. \eqref{para2}. We can note that, in this case, the constant approximation would not be appropriate, since the difference between $\varphi_1(0)$ and $\varphi_1(\pi R)$ is more than $20 \%$. Moreover, $\varphi_1(0) \simeq19$ GeV while $v_1 \simeq85$ GeV, so the corresponding difference is even greater than that of the previous case.

\section{Summary}
In this work we have shown how the explicit breaking of translational
invariance on the extra dimension induced by delta-like interactions 
between scalar bulk and brane fields translates into the vacuum configuration.
This effect modifies
the naively expected pattern of spontaneous symmetry breakdown 
in extra dimensional extensions of the Standard Model containing such terms. 
In particular we have found that, if a general form for the scalar potential is considered, constant non trivial solutions of the equation of motion for 
the scalar fields on the bulk  cannot be found. We are thus forced to consider a 
vacuum configuration for the scalar bulk field that depends on the extra coordinate $y$.

We have used a simple two-Higgs model to illustrate these effects, and, in particular,
we have derived the shape of the vacuum configuration in two examples: in the first one, the $y$ dependence is weak, so that a constant configuration may still be a good approximation; however, the value of the vacuum expectation value on the brane of the scalar bulk field is significantly shifted with respect to the case with no brane-bulk interactions, and this could cause a modification of the Kaluza-Klein spectrum of the bulk fields after the spontaneous symmetry breaking. In the second example, the $y$ dependence is much stronger, and a constant solution would only be a poor approximation to the actual vacuum configuration.

Future developments of this work include the calculation of how the Kaluza-Klein spectrum of both the scalar and gauge fields is modified in a model with brane-bulk interactions, or how these effects modify the scattering of longitudinal gauge bosons among themselves and with Higgs bosons. In addition, one can test how a $y$-dependent vacuum expectation value of the Higgs field would modify the generated fermion masses.

\section*{Acknowledgments}

We thank M. P\'erez-Victoria for useful comments and discussions.
Work partially funded by an INFN-MEC/CICYT Florence-Madrid Collaboration Contract.
For their warm hospitality,
J.R.Pel\'aez thanks the INFN Sezione di Firenze, and F. Coradeschi, D. Dominici, S. De Curtis thank
the Dep. de F\'{\i}sica Te\'orica II of the Complutense University of Madrid; this work was carried out in both places.
J.R.P. research is partially funded by Spanish CICYT contracts
FPA2005-02327, BFM2003-00856 as well as Banco Santander/Complutense
contract PR27/05-13955-BSCH.


\begin{thebibliography}{10}

\bibitem{ArkaniHamed:1999dc}
N.~Arkani-Hamed and M.~Schmaltz,
\newblock Phys. Rev. {\bf D61}, 033005 (2000), [hep-ph/9903417].

\bibitem{Georgi:2000wb}
H.~Georgi, A.~K. Grant and G.~Hailu,
\newblock Phys. Rev. {\bf D63}, 064027 (2001), [hep-ph/0007350].

\bibitem{Kaplan:2001ga}
D.~E. Kaplan and T.~M.~P. Tait,
\newblock JHEP {\bf 11}, 051 (2001), [hep-ph/0110126].

\bibitem{Manton:1988az}
N.~S. Manton and T.~M. Samols,
\newblock Phys. Lett. {\bf B207}, 179 (1988).

\bibitem{Grzadkowski:2004mg}
B.~Grzadkowski and M.~Toharia,
\newblock Nucl. Phys. {\bf B686}, 165 (2004), [hep-ph/0401108].

\bibitem{George:2006gk}
D.~P. George and R.~R. Volkas,
\newblock Phys. Rev. {\bf D75}, 105007 (2007), [hep-ph/0612270].

\bibitem{Davies:2007xr}
R.~Davies, D.~P. George and R.~R. Volkas,
\newblock arXiv:0705.1584 [hep-ph].

\bibitem{Toharia:2007xe}
M.~Toharia and M.~Trodden,
\newblock arXiv:0708.4005 [hep-ph].

\bibitem{Rubakov:1983bb}
V.~A. Rubakov and M.~E. Shaposhnikov,
\newblock Phys. Lett. {\bf B125}, 136 (1983).

\bibitem{Carena:2002me}
M.~S. Carena, T.~M.~P. Tait and C.~E.~M. Wagner,
\newblock Acta Phys. Polon. {\bf B33}, 2355 (2002), [hep-ph/0207056].

\bibitem{Davoudiasl:2002ua}
H.~Davoudiasl, J.~L. Hewett and T.~G. Rizzo,
\newblock Phys. Rev. {\bf D68}, 045002 (2003), [hep-ph/0212279].

\bibitem{delAguila:2003bh}
F.~del Aguila, M.~Perez-Victoria and J.~Santiago,
\newblock JHEP {\bf 02}, 051 (2003), [hep-th/0302023].

\bibitem{delAguila:2003gv}
F.~del Aguila, M.~Perez-Victoria and J.~Santiago,
\newblock Acta Phys. Polon. {\bf B34}, 5511 (2003), [hep-ph/0310353].

\bibitem{delAguila:2003gu}
F.~del Aguila, M.~Perez-Victoria and J.~Santiago,
\newblock Eur. Phys. J. {\bf C33}, s773 (2004), [hep-ph/0310352].

\bibitem{Masip:1999mk}
M.~Masip and A.~Pomarol,
\newblock Phys. Rev. {\bf D60}, 096005 (1999), [hep-ph/9902467].

\bibitem{Casalbuoni:1999ns}
R.~Casalbuoni, S.~De~Curtis, D.~Dominici and R.~Gatto,
\newblock Phys. Lett. {\bf B462}, 48 (1999), [hep-ph/9907355].

\bibitem{Delgado:1999sv}
A.~Delgado, A.~Pomarol and M.~Quiros,
\newblock JHEP {\bf 01}, 030 (2000), [hep-ph/9911252].

\bibitem{Rizzo:1999br}
T.~G. Rizzo and J.~D. Wells,
\newblock Phys. Rev. {\bf D61}, 016007 (2000), [hep-ph/9906234].

\bibitem{Muck:2001yv}
A.~Muck, A.~Pilaftsis and R.~Ruckl,
\newblock Phys. Rev. {\bf D65}, 085037 (2002), [hep-ph/0110391].

\bibitem{DeCurtis:2002nd}
S.~De~Curtis, D.~Dominici and J.~R. Pelaez,
\newblock Phys. Lett. {\bf B554}, 164 (2003), [hep-ph/0211353].

\bibitem{Rubakov:2002fi}
V.~A. Rubakov,
\textit{Classical theory of gauge fields},
\newblock Princeton, USA: Univ. Pr. (2002) 444 p.

\bibitem{Stegun:1964}
M.~Abramowitz and I.~Stegun,
\textit{Handbook of mathematical functions},
\newblock Dover, New York, 1964.

\bibitem{Grad:1964}
I.~S. Gradshtein and I.~M. Ryzhik,
\textit{Tables of Integrals, Series, and Functions},
\newblock Pergamon, New York, 1964.

\end{thebibliography}

\end{document}